\documentclass[twocolumn,showpacs,preprintnumbers,amsmath,amssymb]{revtex4}
\usepackage{graphicx}
\usepackage{dcolumn}
\usepackage{bm}
\usepackage{xcolor}
\begin{document}

\title{Measurement of the Spin Tune Using the Coherent Spin Motion of Polarized Proton in a Storage Ring}

\author{H. Huang,   J. Kewisch, C. Liu, A. Marusic, W. Meng,  F. M\'eot, P. Oddo,  V. Ptitsyn, V. Ranjbar, \\ T. Roser,  and W. B. Schmidke}
\affiliation{Brookhaven National Laboratory, Upton, New York 11973, USA}

\begin{abstract}
This paper reports the first spin tune measurement at high energies (24~GeV and 255~GeV) with a driven
coherent spin motion.
To maintain polarization in a polarized proton collider, it is important to know the spin tune of the polarized proton beam, which is defined as the number of full  spin precessions per revolution.
A nine-magnet spin flipper has demonstrated high spin-flip efficiency  in the presence of two
Siberian snakes~\cite{b_flip}. The spin flipper drives a spin resonance with a given frequency (or tune) and strength. When the drive tune is close to the spin tune, the proton spin direction is not vertical anymore, but precesses around the vertical direction.
By measuring the precession frequency of the horizontal component the spin tune can be precisely measured.
{A driven coherent spin motion and fast turn-by-turn polarization measurement are keys to the measurement.}
The vertical spin direction is restored after turning the spin flipper off and the polarization value is not affected by the measurement. The fact that this manipulation preserves the polarization makes it possible to measure the spin tune during operation of a high energy accelerator.
\end{abstract}

\maketitle
{\it Introduction--}
There have been multiple efforts to understand the origin of the nucleon spin structure  since the discovery of the spin anomaly  by the European Muon Collaboration~\cite{b_ashman}. Colliders with polarized  beams play an important role in  the precision experiments to explain the  proton spin structure~\cite{b_adam}-\cite{b_adare} and also are important topics in accelerator physics.   In this paper, we report the first measurement of driven coherent spin motion in a high energy collider.
These techniques of manipulating polarization properties of beams complement the well established techniques of controlling of orbit properties.

To avoid polarization loss from depolarizing resonances during acceleration and at store, high energy polarized proton colliders require full Siberian snakes, which are specially arranged magnets to rotate the spin by 180$^{\circ}$ around an axis in the horizontal plane~\cite{b_snake}. For the Relativistic Heavy Ion Collider (RHIC), a pair of Siberian snakes are installed in each ring.
This configuration yields a spin tune $\nu_{\rm s}$ of $\frac {1}{2}$~\cite{b_igor},
defined as the number of spin precessions per turn.
A spin tune of $1 \over 2$ avoids all depolarizing resonances as long as the vertical betatron tune $\nu_{\rm y}$ is not also $1 \over 2$.
However, the higher-order   resonances (snake resonances)~\cite{b_snk} still can  lead to polarization loss. The resonance condition is
 \begin{eqnarray}
\nu_{\rm s}=k+m\nu_{\rm y},
\end{eqnarray}
where $k$ and $m$  are integers.
 Snake imperfections and  closed orbit errors  can also shift the spin tune away from $\frac {1}{2}$. This shift leads to a shift of snake resonance locations and  limits  the possible operating parameters of the accelerator~\cite{b_lee}.

 To avoid these higher order resonance conditions, knowing both the vertical betatron tune and the spin tune accurately is important. Betatron tune measurements have been done with various methods in synchrotrons~\cite{b_chao}. It is typically measured with coherent  turn-by-turn beam oscillations in response to a beam excitation.
The spin tune measurement is much harder.
A nine-magnet spin flipper has been used  in RHIC to flip spin~\cite{b_flip}.
This is accomplished by sweeping the drive tune of the spin flipper ac dipole across the spin tune with the proper crossing speed.
The spin flipper can also be used to drive an artificial spin resonance at a fixed tune.
 When the drive tune is near  the spin tune, the polarization direction is moved away from vertical and  is precessing around the vertical direction.  The vertical polarization measurement as a function of drive tune in the vicinity of the expected spin tune can give a direct measurement of the spin tune.
Examples are shown in Fig.~2 of Ref.~\cite{b_flip}. However, such a measurement often results in decoherence of the spin motion and loss of polarization, so it requires several fills of  freshly polarized proton beams. It becomes very time consuming if acceleration to higher energy is needed.

In principle, the spin tune can be measured with a similar idea as the  betatron tune measurement: measuring the spin response to a driven spin coherence. Such a method  can also be  non-destructive. A coherent spin precession around the vertical direction can be adiabatically induced by driving the ac spin rotator at a drive tune near the spin tune.

If the undisturbed stable spin direction is vertical,  the vertical component of polarization $P$ in the neighborhood of an isolated spin resonance
 is given by~\cite{b_bai}:
\begin{eqnarray}
P_{\rm y}=\frac{\nu_{\rm s}-\nu_{\rm osc}} {\sqrt{|\nu_{\rm s}-\nu_{\rm osc}|^2+|\epsilon|^2}} \, ,
\label{b_py}
\end{eqnarray}
where $\epsilon$ is the strength of the  driven spin resonance  and $\nu_{\rm osc}$ is the drive tune. The horizontal component
oscillates with $\nu_{\rm osc}$:
\begin{eqnarray}
P_{\rm x}=\frac{|\epsilon|} {\sqrt{|\nu_{\rm s}-\nu_{\rm osc}|^2+|\epsilon|^2}}\cos(2\pi \nu_{\rm osc} i-\Psi) \, ,
\label{b_px}
\end{eqnarray}
where  $i$ is the $i$th orbital revolution and $\Psi$ is
the initial phase offset. Eqs.~(\ref{b_py}-\ref{b_px}) describe  the  vertical and horizontal components  in a
perfect accelerator in the presence of a single isolated spin resonance.
The ratio of the amplitude part  of Eq.~(\ref{b_px}) $\hat{P_{\rm x}}$ and $P_{\rm y}$ gives the difference between $\nu_{\rm s}$ and $\nu_{\rm osc}$:
\begin{eqnarray}
 \tan \theta_0 = \frac{\hat{P_{\rm x}}}{P_{\rm y}} = \frac{|\epsilon|}{\nu_{\rm s}-\nu_{\rm osc}} \, ,
\label{b_tan}
\end{eqnarray}
where $\theta_0$ is the opening angle  of  the polarization vector.
With known resonance strength $\epsilon$ from the spin flipper and the drive tune $\nu_{\rm osc}$, the spin tune $\nu_{\rm s}$ can be derived from the measured quantity $\tan \theta_0$.

 There are two advantages to this technique.  First,  it is an adiabatic spin manipulation and can preserve the beam polarization. Second, this is a relatively fast measurement.
 Hence, this technique is ideal for measuring the spin  tune at the store energy of a high-energy polarized synchrotron, such as RHIC or a future polarized electron ion collider~\cite{b_accardi}.
 The spin tune measurement with coherent spin  motion has been used
 for deuteron beams~\cite{b_eversmann} at low energy ($\sim$1~GeV) in COSY~\cite{b_maier}, although the  coherent spin motion was not driven. Such a measurement is important for the spin manipulation~\cite{b_hempelmann} needed for the storage-ring-based electric dipole moment measurement\cite{b_orlov}.  In this paper  we report the first spin tune measurement at high energies (24~GeV and 255~GeV) for protons using driven coherent spin motion.

{\it Requirements for the experiment--}
For the success of a spin tune measurement using driven coherent spin motion, several conditions have to be met.
First, a large enough oscillation amplitude needs to be generated. This requires a strong enough driven spin resonance  and a small enough separation between the drive tune and the spin tune.
Second,  the spin tune spread needs to be small. This  allows  a drive tune close to the resonance and significantly enhances the signals.
Third, the measurement requires a  large data sample  from the polarimeter and the polarization must be measured
as a function of the oscillation phase.
 Although the vertical component is constant, the horizontal component is changing from turn to turn. To measure such an oscillation accurately, the polarimeter needs to measure the polarization over many turns as a function of the oscillation phase.

The RHIC spin flipper  can induce a spin resonance with a strength of $\epsilon = 0.00057$. At injection energy, the strength is limited to 0.00024   due to the larger beam size. For this method to work, the separation between $\nu_{\rm s}$ and $\nu_{\rm osc}$ should be similar to the strength to generate a sizable horizontal polarization component.

{\it Polarimeter and spin vector measurements}--
The RHIC p-carbon (pC) polarimeters measure recoil carbon asymmetries~\cite{b_junji} in the
Coulomb-Nuclear Interference (CNI) region~\cite{b_kopeliovich}-\cite{b_trueman}.
Carbon nuclei are scattered in the plane transverse to the polarized proton direction
with an azimuthal angle $\theta$ distribution
\begin{eqnarray}
dN/d\theta \propto 1 + a \sin(\theta - \theta_{\rm s}) \, .
\label{eq-dndtheta}
\end{eqnarray}
Here $a = A_{\rm N} P$,
where $A_{\rm N}$ is the CNI asymmetry for 100\% polarization, which
has values $\sim 10^{-2}$,
and $P$
 at RHIC has values $\sim$ 0.5-0.8.
$\theta_{\rm s}$ is the azimuthal angle of the spin direction in the plane transverse to the beam direction.
The small value of the oscillatory term, $a \sim$ few$\times 10^{-3}$, dictates that
a large statistical sample is necessary to accurately measure it on top of the flat underlying component.

\begin{figure}[ptb]
\includegraphics[width=3.4in, angle=-0]{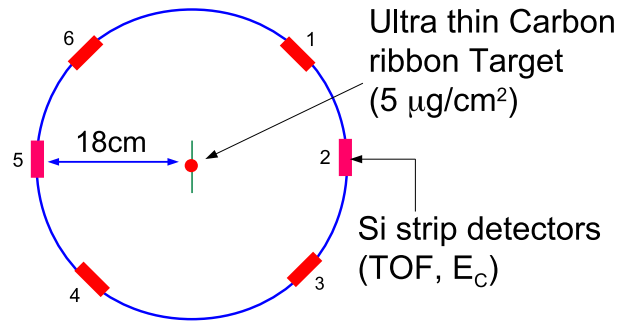}
\caption{Layout of a RHIC p-carbon polarimeter. The target and beam are represented by a vertical line and a dot, respectively.   }
 \label{fig-pClayout}
\end{figure}

Figure~\ref{fig-pClayout} shows a pC polarimeter layout in the plane transverse to the beam direction.
Six silicon strip detectors are arranged as shown approximately 18 cm from the beam; they measure
energy ($\rm E_c$) and time of flight(TOF), allowing selection of the scattered carbon nuclei.
An ultra thin carbon target is moved into the beam during polarimeter measurements~\cite{b_huang}.
The detectors below and above the horizontal plane (Nos. 1, 3, 4 and 6 in Fig.~\ref{fig-pClayout}) make it possible to measure both vertical and horizontal components of beam polarization.

In operation  RHIC is filled with approximately equal numbers of spin up and spin down proton bunches.
The carbon hits are recorded with a unique bunch crossing identifier,
and hits for up and down bunches are counted separately.
A change from spin up to spin down amounts to a shift $\theta_{\rm s} \rightarrow \theta_{\rm s} + \pi$ in Eq.~(\ref{eq-dndtheta}).
The asymmetry of carbon hits, i.e. the relative difference of up and down bunch hits,
normalized by the number of protons on target for each case, as a function of azimuthal angle is then
\begin{eqnarray}
{\rm Aymmetry}|_{\theta} = a \sin(\theta - \theta_{\rm s}) \, .
\label{eq-asymtheta}
\end{eqnarray}

\begin{figure}[ptb]
\includegraphics[width=2.5in, angle=-0]{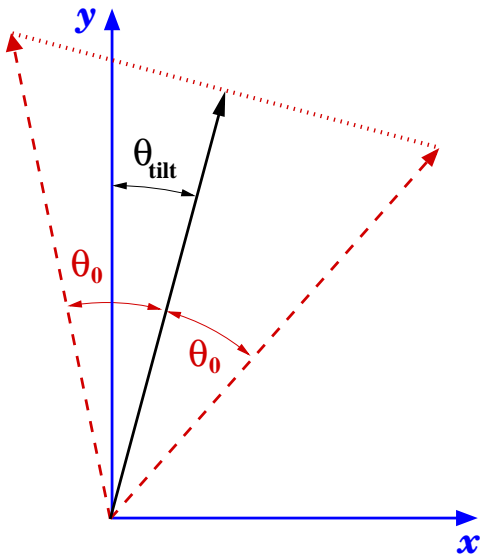}
\caption{Projection of the spin  vector into the transverse plane when the spin tune is
near a spin resonance. The spin  oscillates around the stable spin direction (solid arrow) between the two boundaries (dashed arrows) over many orbit turns. }
 \label{fig-STMparams}
\end{figure}

During normal RHIC operation, the main result from the polarimeters is $a$,
which determines the magnitude of the beam polarization.
  In the present study we focus on $\theta_{\rm s}$, the azimuthal angle of the spin vector in the plane
transverse to the beam direction, and how it is influenced by the coherent spin motion.
{ To measure the driven coherent spin motion, recoil carbon events need to be recorded on a turn-by-turn base.}
Figure~\ref{fig-STMparams} shows the spin precession
projected onto the $x-y$ plane transverse to the beam direction.
The pC polarimeter measures the spin vector projection in this plane.
With driven coherent spin motion the spin vector in this plane
oscillates over the range shown by the two dashed  arrows, with a period equal to
that of the driven resonance.
The amplitude of the precession is $\theta_0$  from Eq.~(\ref{b_tan});
$\theta_{\rm tilt}$  is an arbitrary offset between vertical
and the stable spin direction.
From $P_{\rm x}/P_{\rm y}$
the spin  azimuthal angle  $\theta_{\rm s}$ measured by the pC polarimeter
with a possible tilt angle $\theta_{\rm tilt}$
will follow the precession
\begin{eqnarray}
\frac {P_{\rm x}} {P_{\rm y}}=\tan(\theta_{\rm s} - \theta_{\rm tilt}) = \tan \theta_0 \cdot \cos(2\pi \nu_{\rm osc}i - \Psi) \, .
\label{eq-thetaprecess}
\end{eqnarray}
Note that only  the two transverse components of the polarization can be measured. If the spin direction has  a significant longitudinal component in addition to the angle $\theta_{\rm tilt}$, the simple form of Eq. (\ref{eq-thetaprecess}) should be modified.

{\it Spin tune measurement results--}
The  experiment was carried out at two different energies, injection at 24~GeV and store at 255~GeV.
The revolution frequency in RHIC is about 78.20~kHz.
The bunch pattern was 120 bunches in the ring and
RHIC bunch crossings were used as a clock signal for the analysis.
For these measurements, a signal from the resonance drive was provided to the
polarimeter readout, which allowed alignment of the phase of carbon hits within
one period of the resonance drive.
The drive signal was read with an accuracy of two bunch crossings, whereas
the typical period of the drive was $\sim 240$ bunch crossings (for a drive tune near 0.5), so the phase of
carbon hits was known to within 1\% of a period.

 In the experiment, the spin tune was first roughly located by sweeping the drive tune.  If the polarization sign flips, the spin tune is covered by the sweep range. This method successively divides the drive tune sweep range in half over the intervals that exhibit a spin-flip. The progressively narrowing tune sweep ranges caused increased polarization loss and eventually converges on a range.  This typically gives a narrow range for the spin tune location. Then, with fresh beam, the drive was turned on adiabatically at fixed tune and the driven coherent spin motion was measured
 with the polarimeter. After the measurement, the drive was turned off adiabatically and the original spin direction was restored.

\begin{figure}[ptb]
\includegraphics[width=3.4in, angle=-0]{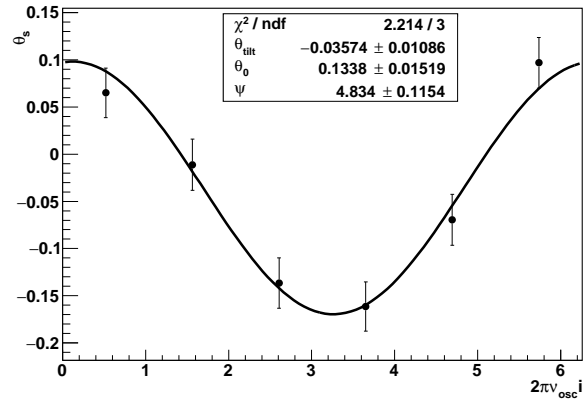}
\caption{Measured spin  azimuthal angle as a function of driven oscillation phase at 24~GeV with drive tune as 0.498. All angles ($\theta_{\rm s}, \theta_{\rm tilt}, \theta_0, \Psi$) are in the unit of radian. The non-zero $\theta_{\rm tilt}$ means that  the stable spin direction is tilted  away from vertical. }
 \label{fig-STMset2}
\end{figure}

Figure~\ref{fig-STMset2}  shows
$\theta_{\rm s}$ versus one cycle of drive phase for one drive setting.
For statistical accuracy, the carbon hits were grouped in 6 bins of 40 bunch crossings,
spanning nearly one entire  drive cycle; the mean spin  azimuthal angle $\theta_{\rm s}$ was measured
in each bin.
The curve is fit to the function, from rearrangement of Eq.~(\ref{eq-thetaprecess}):
\begin{eqnarray}
\theta_{\rm s}(i) = \theta_{\rm tilt} + \tan^{-1} \left( \tan \theta_0 \cdot \cos(2\pi\nu_{\rm osc}i - \Psi) \right) \, .
\label{eq-thetafit}
\end{eqnarray}
The arbitrary phase offset $\Psi$ depends on the propagation time of proton bunches
from the drive  to the polarimeter, and the cable delay of the signal from the drive to the polarimeter readout.

\begin{table}
\begin{tabular}{ccccc}
\hline
set  &   $\theta_0$ (rad)      &     $\nu_{\rm osc}$  &   $\nu_{\rm s}$ from coherence     &   $\nu_{\rm s}$ from flip \\ \hline
1&  0.273$\pm$0.059      &   0.499        &      0.4999$\pm$0.0002    &     0.4975-0.5 \\
2&   0.134$\pm$0.015     &   0.498        &      0.4998$\pm$0.0002    &     0.4975-0.5 \\
3&  0.109$\pm$0.015    &   0.5004        &    0.5026$\pm$0.0003      &   0.5022-0.5025 \\
4&  0.132$\pm$0.021     &   0.5009        &    0.5027$\pm$0.0003      &   0.5022-0.5025 \\
5&   0.062$\pm$0.015   &   0.499       &       0.4951$\pm$0.0010    &       0.491-0.495 \\
6&  0.263$\pm$0.033     &    0.494        &      0.4961$\pm$0.0003    &     0.495-0.4965 \\
7&  0.174$\pm$0.024     &    0.493      &       0.4962$\pm$0.0005   &      0.495-0.4965 \\ \hline
\end{tabular}
\caption{Spin tune measurement results. The first five cases are at 24~GeV and the last two cases are at 255~GeV.  The precession amplitude angle $\theta_0$ is in the second column. The third column is the drive tune of the ac dipoles. The derived spin tune from driven coherence is  in the forth column. Last column is the spin tune range from spin flipper operation. }
\label{tab:a}
\end{table}

 The measured $\theta_0$ and derived spin tune for 7 sets of measurements are shown in Table~\ref{tab:a}. The separation of the drive tune and the spin tune varies  from 0.001 to 0.004.
For each of the three pairs of measurements done under the same conditions, the results are consistent with each other within the statistical errors.
Within statistical errors, the derived spin tune from the driven spin coherence (column 4 of Table~\ref{tab:a}) agrees with the range obtained from the spin-flip method (column 5 of Table~\ref{tab:a}).

Among the seven spin tune measurements, five are at injection. The first pair of spin tune measurements were done
with two different drive tunes.
The experimental data and fit  from the second set are shown in Fig.~\ref{fig-STMset2}.
The spin tune was above 0.5 for the next pair of spin tune measurements (sets 3 and 4 in Table~\ref{tab:a} ).
Although the spin tune range  determined with spin-flip method was fairly small (0.5022-0.5025), the spin tune from coherent spin motion measurements of the pair are self-consistent with each other and they
are in agreement with the results from the spin-flip method within statistical errors.

\begin{figure}[ptb]
\includegraphics[width=3.4in, angle=-0]{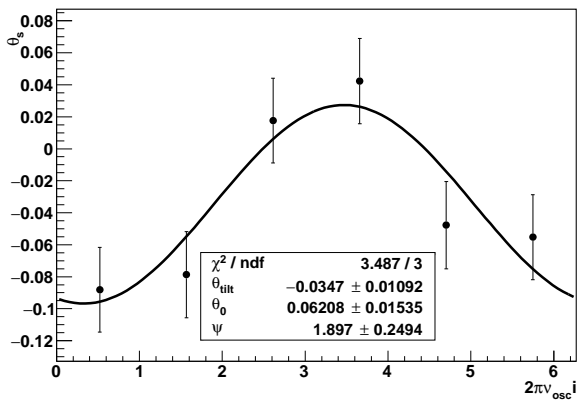}
\caption{Measured spin  azimuthal  angle as a function of driven oscillation phase at 24~GeV with drive tune as 0.499.  }
 \label{fig-STMset5}
\end{figure}

\begin{figure}[ptb]
\includegraphics[width=3.4in, angle=-0]{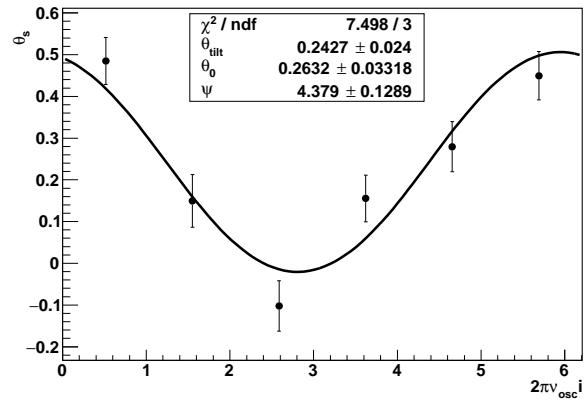}
\caption{Measured spin azimuthal angle as a function of driven oscillation phase at 255~GeV with drive tune as 0.494.}
 \label{fig-STMset6}
\end{figure}

In the 5th set of spin tune measurement, the Siberian snakes were tuned to move the spin tune far away from 0.5, and the spin-flip method gave the spin tune between 0.491-0.495. The drive tune for the spin coherence measurement was 0.499.
The experimental data and fit are shown in Fig.~\ref{fig-STMset5}. Due to larger tune separation, the asymmetry oscillation amplitude was smaller and the relative statistical uncertainty was larger.
It should be noted that the drive tune was higher than the spin tune in this case, and the phase of the coherent spin motion also changed by $\pi$, or flipped sign. Since the phase of the spin coherence motion depends on whether the drive tune is above or below the spin tune, the spin tune is determined, not just its distance to the drive tune.

The last pair of spin coherence measurements were done at 255~GeV.
 The 6th set of  the experimental data and fit are shown in Fig.~\ref{fig-STMset6}.
The spin tune was also determined as between 0.496 and 0.4965 by a fixed drive tune scan (see Fig.~2 of Ref.{\cite{b_flip}).
Note that there is a quite large tilt angle of the spin of 0.25 rad at the store energy~\cite{b_vahid}. The tilt angle is probably due to a nearby orbit bump separating the two RHIC beams~\cite{b_meot}. More study is needed to quantify this effect. Even with such a tilt angle, the spin tune still can be measured with a driven spin coherence.
The spin tune measured from coherent spin motion
 is in agreement with the results from the spin-flip and fixed drive tune methods.
%This implies that Eq.(\ref{eq-thetaprecess}) is valid  for RHIC and there is no significant   longitudinal  component of the stable spin direction.

If there is a longitudinal component of the stable spin direction,  %Eq.(\ref{eq-thetaprecess}) should be modified. In this case,
the stable spin direction has an angle of $\theta_{\ell}$  with respect to the $(x,y)$ plane.
For a  rotation around the $x$-axis, the transverse polarization ratio  becomes
 \begin{eqnarray}
%\frac {P_{\rm x}} {P_{\rm y}}=
\tan(\theta_{\rm s} - \theta_{\rm tilt}) ={{ \tan \theta_0  \cos\phi} \over {
 \cos \theta_{\ell}+\sin \theta_{\ell}\tan \theta_0 \sin\phi}}
\label{eq-long}
\end{eqnarray}
where $\phi=2\pi \nu_{\rm osc}i - \Psi$.
Expanding the function to the first order of $\theta_{\ell}$,
\begin{eqnarray}
\tan(\theta_{\rm s} - \theta_{\rm tilt})&  \approx & {\tan \theta_0  \cos\phi}
 (1-\theta_{\ell} \tan \theta_0\sin\phi )   \nonumber \\
 & =&   \tan \theta_0 \cos\phi - {{\theta_{\ell}} \over {2}} \tan^2 \theta_0 \sin 2\phi
\end{eqnarray}
The second term is in the 2nd harmonic of the phase angle. From the shapes of the data points in Figs.~\ref{fig-STMset2}-\ref{fig-STMset6}, this component is small. In addition, for the second term to be small, $\theta_{\ell}$ must satisfy
$\theta_{\ell}\tan \theta_0 \ll{1}$.
The  $\theta_{\ell}$ from the   data fits are consistent with zero within one sigma of the statistical error. The largest case is $\theta_{\ell}=0.06\pm 0.06$ rad while $\theta_0$ is in the range of 0.06 to 0.27, which satisfies the constraint. The resulting difference in $\theta_0$ with such  a small $\theta_{\ell}$  is in the order of $10^{-4}$ and the effect on spin tune is negligible.
%, at a $10^{-5}$ level.

{\it Conclusion--}
Driven coherent spin motion has been used to measure the spin tune in RHIC at 24~GeV and 255~GeV.
The results show that the spin tune can be measured by driven spin coherence when the tune separation is small enough.  For it to work, the drive tune needs to be close to the spin tune, which requires a  small spin tune spread. In RHIC, where a pair of Siberian snakes are used, the small spin tune spread was achieved by the reduction of dispersion slope difference at  the two Siberian snakes~\cite{b_liu, b_vadim}.
These experimental results prove that it is possible to routinely measure the spin tune of polarized  proton beams---the most important polarized beam parameter---which will lead to more stable and optimized operation of a high-energy polarized collider, such as
RHIC or a future polarized electron ion collider.
\acknowledgments
{The work was supported by many in the Collider Accelerator  Department at Brookhaven National Laboratory.
We  thank   M. Bai and A. Poblaguev for the early work of the driven spin coherence in RHIC.
I. Alekseev, Z. Chang and D. Svirida are acknowledged for their   assistance in polarimetry measurements.
Work was supported by Brookhaven Science Associates, LLC, under Contract  No.  DE-AC02-98CH10886  with  the  U.S. Department of Energy.}


\begin{thebibliography}{99}
\bibitem{b_flip} H. Huang, {\it et al.},  Phys. Rev. Lett.
{\bf 120}, 264804(2018).
\bibitem{b_ashman} J. Ashman, {\it et al.},  Phys. Lett.
{\bf B206}, 304(1988).
\bibitem{b_adam} J. Adam, {\it et al.},  Phys. Rev.
{\bf 98}, 032011(2018).
\bibitem{b_adare} A. Adare, {\it et al.},  Phys. Rev.
{\bf 94}, 112008(2016).
\bibitem{b_snake} Ya. S. Derbenev and A. M. Kondratenko, Part. Accel. {\bf 8},
115 (1978).
\bibitem{b_igor} I. Alekseev, {\it et al.}, Nucl. Instrum. Methods Phys. Res.,    {\bf A499}, 392 (2003).
\bibitem{b_snk}  S.Y. Lee and S. Tepikian, Phys. Rev. Lett. 56 (1986) 1635.
\bibitem{b_lee}  S.Y. Lee, {\it Spin Dynamics and Snakes in Synchrotrons}, (World Scientific, Singapore, 1997).
\bibitem{b_chao} A.W. Chao and M. Tigner, {\it Handbook of Accelerator Physics and Engineering}, (World Scientific, Singapore, 1999).
\bibitem{b_bai} M. Bai and T. Roser,  Phys. Rev. ST AB {\bf 11}, 091001(2008).
\bibitem{b_accardi} A. Accardi, {\it et al.}, Eur. Phys. J. A {\bf 52}, 268(2016).
\bibitem{b_eversmann} D. Eversmann, {\it et al.}, Phys. Rev. Lett. {\bf 115}, 094801(2015).
\bibitem{b_maier} R. Maier, Nucl. Instrum. Methods Phys. Res., Sect. A
{\bf 390},1(1997).
\bibitem{b_hempelmann} N. Hempelmann, {\it et al.}, Phys. Rev. Lett. {\bf 119}, 014801(2017).
\bibitem{b_orlov} Y.F. Orlov, W. M. Morse, and Y. K. Semertzidis,
Phys. Rev. Lett. {\bf 96}, 214802 (2006).
\bibitem{b_junji} J. Tojo, I. Alekseev, M. Bai, B. Bassalleck, G. Bunce, A. Deshpande, J. Doskow, S. Eilerts, D.E. Fields, {\it et al.}, Phys. Rev. Lett. {\bf 89}, 052302(2002).
\bibitem{b_kopeliovich} B. Z. Kopeliovich and T. L. Trueman,  Phys. Rev. {\bf D64}, 034004(2001).
\bibitem{b_buttimore} N. H. Buttimore, E. Leader  and T. L. Trueman,  Phys. Rev. {\bf D64}, 094021(2001).
\bibitem{b_trueman}  T. L. Trueman,  Phys. Rev. {\bf D77}, 054005(2008).
\bibitem{b_huang} H. Huang and K. Kurita,   AIP Proc.{\bf 868},  3(2006).
\bibitem{b_vahid} V. Ranjbar,  {\it et al.}, {\it RHIC Polarized Proton Operation for 2017}, Proc. of  IPAC 2017, 2190(2017).
\bibitem{b_meot} F. M\'eot,  {\it et al.}, {\it Revisiting RHIC Snakes OPERA Fields N0 Dance}, BNL C-AD Tech Notes AP 590, September 2017(https://www.bnl.gov/isd/documents/95364.pdf).
\bibitem{b_vadim} V. Ptitsyn,  M. Bai, T. Roser, {\it Spin Tune Dependence on Closed Orbit in RHIC}, Proc. of  IPAC 2010, 4641(2010).
\bibitem{b_liu} C. Liu , J. Kewisch, H. Huang, {\it Minimization of Spin Tune Spread by Matching Dispersion Primse at RHIC},  BNL C-AD Tech Notes AP 593, August 2017(https://www.bnl.gov/isd/documents/95271.pdf).
\end{thebibliography}
\end{document}